\documentclass[12pt]{article}
\usepackage{amsmath}
\usepackage{amssymb}
\usepackage{graphicx}
\usepackage{axodraw}
\usepackage{epsfig}
\usepackage{url}
\usepackage{fancybox}
\usepackage[small]{caption}
\begin{document}
\baselineskip 0.6cm

\def\simgt{\mathrel{\lower2.5pt\vbox{\lineskip=0pt\baselineskip=0pt
           \hbox{$>$}\hbox{$\sim$}}}}
\def\simlt{\mathrel{\lower2.5pt\vbox{\lineskip=0pt\baselineskip=0pt
           \hbox{$<$}\hbox{$\sim$}}}}
\def\simprop{\mathrel{\lower3.0pt\vbox{\lineskip=1.0pt\baselineskip=0pt
             \hbox{$\propto$}\hbox{$\sim$}}}}
\def\bra#1{\left< #1 \right|}
\def\ket#1{\left| #1 \right>}
\def\inner#1#2{\left< #1 | #2 \right>}

\begin{titlepage}

\begin{flushright}
UCB-PTH-12/04 \\
\end{flushright}

\vskip 1.3cm

\begin{center}
{\Large \bf Quantum Mechanics, Gravity, and the Multiverse}

\vskip 0.7cm

{\large Yasunori Nomura}

\vskip 0.4cm

{\it Berkeley Center for Theoretical Physics, Department of Physics,\\
 University of California, Berkeley, CA 94720, USA}

\vskip 0.1cm

{\it Theoretical Physics Group, Lawrence Berkeley National Laboratory,
 CA 94720, USA}

\vskip 0.8cm

\abstract{The discovery of accelerating expansion of the universe has 
led us to take the dramatic view that our universe may be one of the many 
universes in which low energy physical laws take different forms:\ the 
multiverse.  I explain why/how this view is supported both observationally 
and theoretically, especially by string theory and eternal inflation. 
I then describe how quantum mechanics plays a crucial role in understanding 
the multiverse, even at the largest distance scales.  The resulting 
picture leads to a revolutionary change of our view of spacetime and 
gravity, and completely unifies the paradigm of the eternally inflating 
multiverse with the many worlds interpretation of quantum mechanics. 
The picture also provides a solution to a long-standing problem in 
eternal inflation, called the measure problem, which I briefly describe.}

\end{center}
\end{titlepage}

\section{Introduction---Why the Multiverse}
\label{sec:intro}

Why does our universe have the structure we see today?  For example, 
why do the quarks and leptons have the observed masses, and why are 
there four elementary (electromagnetic, weak, strong, and gravitational) 
forces acting on them?  At some point in the history of elementary 
particle physics, we hoped that all these questions would be answered 
once we had figured out the ``fundamental theory of nature.''  Namely, 
mathematical consistency of the ultimate theory would not allow any 
other world than the one we see today.  In the past few decades, however, 
we have gradually been asked---or forced---to consider that this may not 
be the case:\ many (if not all) of the structure we observe are due 
to our very own existence in the huge {\it multiverse}, a collection 
of many different universes in which everyday physical laws take 
different forms.

A shocking revelation which has hugely impacted our thinking came in 
1998 when it was discovered that the expansion of the universe is 
accelerating~\cite{Riess:1998cb}.  Because the gravitational force 
between two bodies is attractive, the expansion of the universe must 
be decelerating if it contains only matter (in any form, even dark 
matter).  In order to explain the peculiar phenomenon of accelerating 
expansion, the universe must be filled with energy with ``negative 
pressure'' (called dark energy).  The simplest possibility for such 
a strange entity is energy of the vacuum:\ the observed acceleration 
is accounted for if the vacuum has energy density
\begin{equation}
  \rho_\Lambda \sim 7 \times 10^{-30}~{\rm g/cm^3},
\label{eq:rho_Lambda}
\end{equation}
which is comparable to the average energy density of matter in the 
universe, $\rho_{\rm matter} \sim 3 \times 10^{-30}~{\rm g/cm^3}$. 
A question is why these two totally different entities (matter and 
vacuum!) are so close in density {\it in the current universe}. 
This is very mysterious, especially given that even time dependence 
of the two quantities $\rho_{\rm matter}$ and $\rho_\Lambda$ are 
different: $\rho_{\rm matter} \sim 1/t^2$ and $\rho_\Lambda \sim 
{\rm const}.$

In fact, a theoretical estimate of the energy density of the vacuum 
has been a notoriously difficult problem.  Quantum mechanical 
``corrections'' to the vacuum energy are huge---at least about 60 
orders of magnitude larger than the size allowed by observation. 
This problem has been known as the cosmological constant problem, 
and despite many attempts, it has abhorred a simple theoretical 
solution~\cite{Weinberg:1988cp}.  Until recently, many theorists had 
still been hoping that an yet unknown mechanism will set the vacuum 
energy to be zero, $\rho_\Lambda = 0$, but the discovery of nonzero value 
in Eq.~(\ref{eq:rho_Lambda}) destroyed this hope.  How can the theory 
know when we---the human species---evolve to the point making cosmological 
observations, and set the vacuum energy density close to the matter 
energy density {\it at that particular moment} in the history of 
the universe?

Already back in 1980's, Steven Weinberg realized the difficulty of 
solving the problem, and considered the possibility that the origin 
of the smallness of the vacuum energy might be ``environmental'':\ 
we simply cannot exist if the vacuum energy were (much) larger than 
the matter energy density {\it at the time when relevant structures 
of the universe, such as large galaxies, form} (which is only within 
a few orders of magnitude of the timescale of the evolution of the 
human species)~\cite{Weinberg:1987dv}.  Suppose there are many 
``universes,'' or large enough spacetime regions, in which the vacuum 
energy, $\rho_\Lambda$, takes different values.  Then simple calculations 
can show that unless $|\rho_\Lambda|$ is within a few orders of magnitude 
of $\rho_{\rm matter}$ in the current universe (the universe the human 
species observes), no galaxies, and thus presumably no intellectual 
observers, form.  A prediction of this framework is that, unlike many 
other attempts trying to achieve $\rho_\Lambda = 0$, {\it we expect 
to see nonzero $\rho_\Lambda$} since values of $|\rho_\Lambda|$ much 
smaller than needed for the existence of life are unnatural.  In fact, 
this is what happened in 1998:\ we discovered accelerating expansion 
of the universe which can be caused by the vacuum energy density, 
Eq.~(\ref{eq:rho_Lambda}), that is not much different from the matter 
energy density in magnitude when intellectual life observed the cosmos.

The assumption of multiple universes may seem too big to swallow 
based on a single observation of accelerating expansion of the universe 
(although this is completely consistent with what we have learned 
throughout our history:\ our Earth turned out to be only one of several 
planets in the solar system, which is only one of many such systems 
in the galaxy, which is again one of many in our local cluster, etc). 
If we look at the structure of the theory of elementary particle 
physics and cosmology, however, there are also many ``miracles'' 
that seem to be too good for our own existence; for example, only 
a slight change of certain parameters of the theory seems to lead 
to a completely sterile world, e.g., that without any interesting 
chemistry~\cite{Barrow-Tipler}.  With the multiverse, these apparent 
``miracles'' have a simple explanation---there are many universes 
within the multiverse in which physical properties including the value 
of $\rho_\Lambda$ are different; and only in those universes in which 
the conditions are friendly enough for life, an intellectual observer 
would evolve.  Therefore, there is no surprise if the observer finds 
the structure of physical laws to be ``tuned'' too good for him/her; 
otherwise, he/she is simply not there.

Interestingly, the existence of the multiverse has been suggested 
by theories of elementary particle physics and cosmology.  String 
theory---widely considered to be the best candidate for the ultimate 
theory of nature---predicts the existence of six extra spatial dimensions 
beyond the three we experience in our everyday life~\cite{string-theory}. 
In the old days, people viewed this as a nuisance.  They simply ``hid'' 
these dimensions by postulating that they are too small to see, 
analogous to the direction on a surface of a thin wire perpendicular 
to the direction of the extension.  These extra dimensions, however, 
turned out to be a blessing, rather than a nuisance---because the 
six small dimensions can have a variety of meta-stable configurations, 
string theory can lead to a variety of four (3 spatial $+$ 1 time) 
dimensional theories at our length scales, whose properties---including 
$\rho_\Lambda$---depend on the shape and size of the compactified six 
dimensional space~\cite{Bousso:2000xa}.  This plethora of possible 
different worlds is called the string landscape, and the number of 
such possible worlds is indeed huge:\ people's estimates vary but 
typically give numbers like $O(10^{1000})$.  Moreover, once one 
meta-stable configuration with $\rho_\Lambda > 0$ is realized, 
then exponential expansion of space, called inflation, occurs. 
And it has been known from the 1980's that inflation is generically 
future-eternal~\cite{Guth:1982pn}:\ once it occurs, space keeps 
expanding forever.  Again, some people viewed this as an undesired 
feature, but in the context of the string landscape, it implies that 
all different four dimensional worlds are indeed physically realized 
in spacetime, producing the multiverse.  The way it works is the 
following:\ because of infinite space available, all kinds of ``bubbles'' 
having different properties inside are formed in eternally inflating 
spacetime~\cite{Coleman:1980aw}, and each of these bubbles corresponds 
to a universe with definite physical laws.  It is quite suggestive 
that phenomena many (though not all) people found unwanted, but 
nevertheless indicated by theory, are exactly the elements needed 
to realize the multiverse, and hence to solve the cosmological 
constant problem.

Despite all the good features described above, however, understanding 
the multiverse in eternally inflating spacetime has been notoriously 
difficult because of the infinity introduced by the eternal nature of 
inflation.  In this article, I explain this problem---often called the 
measure problem in eternal inflation~\cite{Guth:2000ka}---and report 
recent progress on this issue:\ quantum mechanics is crucial in 
understanding the multiverse correctly {\it even at the largest 
distance scales}~\cite{Nomura:2011dt,Nomura:2011rb}.  This leads 
to a dramatic change of our view of spacetime and gravity, consistently 
with what we learned about quantum gravity in the past two decades:\ 
the holographic principle~\cite{'tHooft:1993gx} and black hole 
complementarity~\cite{Susskind:1993if}.  We will find that this new 
framework completely unifies the eternally inflating multiverse and 
the many worlds interpretation of quantum mechanics:\ these are absolutely 
the same concept~\cite{Nomura:2011dt}.  We will also see that the notion 
of spacetime is ``reference frame dependent''~\cite{Nomura:2011rb}, 
precisely analogous to that of simultaneity in special relativity.

\section{Predictivity Crisis in Eternal Inflation}
\label{sec:pred-crisis}

The heart of the problem in eternal inflation is well summarized in 
the following sentence by Alan Guth~\cite{Guth:2000ka}:\ ``In an eternally 
inflating universe, anything that can happen will happen; in fact, it 
will happen an infinite number of times.''  Suppose we want to calculate 
the relative probability for events $A$ and $B$ to happen.  Following the 
standard notion of probability, we might define it as the ratio of the 
numbers of times events $A$ and $B$ to happen throughout the whole spacetime
\begin{equation}
  P = \frac{N_A}{N_B}.
\label{eq:P}
\end{equation}
The eternal nature of inflation, however, makes both $A$ and $B$ 
occur infinitely many times: $N_A, N_B = \infty$.  The expression in 
Eq.~(\ref{eq:P}), therefore, is ill-defined.  It seems that we need 
to ``regularize'' spacetime to make both $N_{A,B}$ finite, at least 
at a middle stage of the calculation.

\begin{figure}[t]
\begin{center}
  \includegraphics[width=16cm]{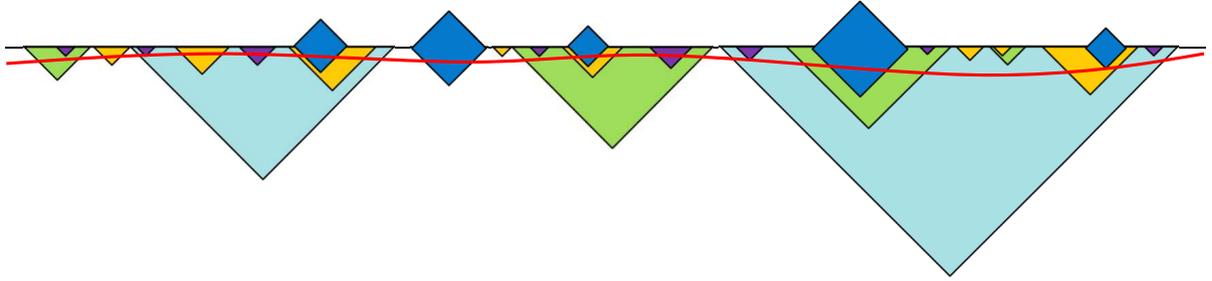}
\caption{A schematic depiction of the eternally inflating multiverse. 
 The horizontal and vertical directions correspond to spatial and 
 time directions, respectively, rescaled such that the propagation 
 of light is always in a $45^\circ$ direction.  Various regions with 
 the inverted triangle or argyle shape represent different universes, 
 which form in other, parent regions through bubble nucleation 
 processes.  While regions closer to the upper edge of the diagram 
 look smaller, it is an artifact of the rescaling made to fit the 
 infinitely large spacetime into a finite drawing---the fractal 
 structure near the upper edge actually corresponds to an infinite 
 number of large (in fact, infinitely large) universes.  A fictitious 
 time cutoff, $t = t_c$, is depicted by a red, curved line.  The 
 number of universes below this line is finite if we focus on an 
 initially finite spatial region.}
\label{fig:cutoff}
\end{center}
\end{figure}
An obvious way to do this is to consider an equal-time cutoff, $t = t_c$, 
and count only events that occur before this cutoff.  Suppose we focus 
only on some finite spatial region at the beginning.  Then, since the 
numbers of events become infinity only due to those that happen in the 
infinite future, the cutoff makes $N_{A,B}$, and hence $P$, finite; 
see Fig.~\ref{fig:cutoff} for a schematic depiction.  We can then 
imagine removing this cutoff by sending $t_c \rightarrow \infty$, 
and obtain a well-defined answer for $P$.  A problem of this procedure 
is that the definition of ``equal time'' is arbitrary.  Already in 
special relativity the concept of equal time depends on an observer, 
but the situation is much worse in general relativity---there is no 
way of uniquely introducing the concept of equal time (even an observer 
dependent one!) for points separated beyond the horizon, especially 
if the system does not possess any obvious symmetry, which is the case 
in the eternally inflating multiverse.  Indeed, one can show that by 
carefully devising the cutoff ``hypersurface'' (a surface of equal time 
in spacetime), we can obtain any value of $P$ we want---the probability 
is determined by how we regularize spacetime!

This extreme sensitivity of predictions on the regularization procedure 
is called the measure problem in eternal inflation.  In fact, the problem 
is much more robust than one might naively think.  Suppose there is 
a meta-stable universe with $\rho_\Lambda > 0$ (more precisely, a 
meta-stable minimum in the space of quantum fields that has positive 
potential energy).  The conventional wisdom says that if $\rho_\Lambda$ 
is smaller than the Planck energy density, $\rho_{\rm Pl} \simeq 5.1 
\times 10^{93}~{\rm g/cm^3}$, then the result of general relativity is 
applicable, so that if the decay rate of such a meta-stable state is 
small enough, it leads to eternal inflation.  This is enough to encounter 
the problem of predictivity described above---it has nothing to do with 
the string landscape, the beginning of the universe, or anything like 
that; in particular, the problem occurs already in a regime where 
quantum gravitational effects have been {\it believed to be} unimportant, 
$\rho_\Lambda \ll \rho_{\rm Pl}$.  This, of course, does not mean that 
such a belief, i.e.\ that the solution to the problem does not involve 
quantum gravity, is correct.  In fact, we will see that the solution 
to the measure problem is intrinsically quantum gravitational.

Another important aspect of the measure problem is that the simplest 
attempt based on semi-classical intuition completely fails.  Imagine 
that at some early time the entire universe was in an inflating phase, 
and that fictitious clocks located at various places are synchronized 
according to natural time. (Because the inflationary phase has 
a large symmetry, such natural time---the flat slicing in technical 
terminology---can be defined.)  Now, we can define a cutoff hypersurface 
as the one on which all the fictitious clocks show the same time, 
and calculate the probability~\cite{Linde:1993xx}.  In particular, 
we can calculate the relative probability of us observing a universe 
with $3{\rm K}$ cosmic microwave background (CMB) to that with 
$2.725{\rm K}$ CMB, which gives
\begin{equation}
  \frac{N_{T_{\rm CMB} = 3{\rm K}}}{N_{T_{\rm CMB} = 2.725{\rm K}}} 
  \sim 10^{10^{59}}.
\end{equation}
Namely, the probability of us seeing a $3{\rm K}$ universe is much, much 
higher than that of seeing a $2.725{\rm K}$ universe as we do!  This 
ridiculous conclusion, called the youngness paradox, arises because space 
expands exponentially in an inflating phase, proportional to $\exp(3Ht)$ 
with $H^{-1}$ being a microscopic timescale, and the rate of creating 
universes like our own in such space is constant {\it per unit physical 
volume per unit time}.  Therefore, the number of universes created at 
later times increases like crazy, hence giving huge bias towards younger 
universes when counted at a fixed time defined through the fictitious 
clocks as described above.

While many proposals have been put forward to solve this and other 
problems, especially by modifying the way to define the cutoff, they 
all look rather ad hoc~\cite{Guth:2000ka}.  Indeed, it is extremely 
uncomfortable that we need to specify the exact way of regulating 
spacetime {\it to define the theory}, beyond the basic principles of 
quantum mechanics and relativity.  It seems that something crucial 
is missing in a way the problem is considered.

\section{The Quantum Multiverse}
\label{sec:multiverse}

We now argue that the missing ingredient is quantum mechanics.  At 
first sight, this statement sounds trivial---since the process of vacuum 
decay (a process creating a universe in another universe through a bubble 
nucleation; see Fig.~\ref{fig:cutoff}) is probabilistic in the usual 
quantum mechanical sense, the entire system must ultimately be treated 
using quantum mechanics.  A surprising thing is that it affects our 
thinking of what spacetime actually is---and hence what the multiverse 
is---{\it at distance scales much larger than the Planck length} $l_P 
\simeq 1.6 \times 10^{-35}~{\rm m}$, conventionally thought to be the 
scale only below which quantum gravitational effects become important.

The basic principle we adopt is that {\it the laws of quantum mechanics 
are not violated when an appropriate description of physics is 
adopted}---from the shortest to the largest scales we ever consider. 
Given the extreme successes of quantum mechanics over the last century, 
this seems to be a reasonable, and in a sense conservative, hypothesis 
to take.  Then the situation of the eternally inflating multiverse does 
not seem much different from those in any usual experiments.  Suppose 
we scatter an electron with a positron, which leads probabilistically 
to different final states: $e^+e^-$, $\mu^+\mu^-$, $e^+e^-e^+e^-$, 
$\cdots$.  One might view this as the initial state $\ket{e^+e^-}$ 
evolving probabilistically into different final states, but this is 
not true.  Since the Schr\"{o}dinger equation is deterministic, the initial 
$\ket{e^+e^-}$ state simply evolves deterministically into some final 
state $\Psi(t=+\infty)$ which, after being decomposed into eigenstates 
of particle numbers, contains many components:
\begin{equation}
  \Psi(t = -\infty) = \ket{e^+ e^-}
\quad\rightarrow\quad
  \Psi(t = +\infty) = c_e \ket{e^+ e^-} + c_\mu \ket{\mu^+ \mu^-} + \cdots,
\label{eq:QFT-evolution}
\end{equation}
where $c_e, c_\mu, \ldots$ are coefficients calculable according to 
the Schr\"{o}dinger equation.  The situation for the multiverse must 
be similar.  Starting from a state corresponding to eternally inflating 
space $\ket{\Sigma}$ at $t = t_0$, it evolves deterministically into 
some state $\Psi(t)$ at time $t$ which, after being decomposed into 
states having well-defined semi-classical spatial geometries, contains 
many components:
\begin{equation}
  \Psi(t = t_0) = \ket{\Sigma}
\quad\rightarrow\quad
  \Psi(t) = \sum_i c_i(t) \ket{\mbox{(cosmic) configuration $i$}},
\label{eq:multiverse-evolution}
\end{equation}
where the absolute value squared of coefficient $c_i(t)$ should give 
the probability of finding the universe in cosmic configuration $i$ 
at time $t$.

Formulating the multiverse in the form of Eq.~(\ref{eq:multiverse-evolution}), 
however, does not solve any of the problem by itself.  What is actually 
the ``multiverse state'' $\Psi(t)$?  To define a quantum state we need to 
specify an equal-time hypersurface on which the state is defined, and there 
is an intrinsic ambiguity in doing this for spatial points separated beyond 
the horizon.  Moreover, even if we follow a region whose spatial extent 
was initially finite, such a region will grow into an infinitely large 
spatial region in which an infinite number of observers will arise, so the 
problem of infinity does persist.  We will find below that when the system 
is treated correctly, the final picture turns out, in fact, like that given 
in Eq.~(\ref{eq:multiverse-evolution})~\cite{Nomura:2011dt,Nomura:2011rb}. 
To see this, however, we need to understand better quantum mechanics in 
a system with gravity, which requires a dramatic revision of our view 
of spacetime.

\subsection{Quantum mechanics in a system with gravity}
\label{subsec:QM-gravity}

Black holes provide important ``laboratories'' to test strong gravitational 
physics.  In 1976, Stephen Hawking found a strange phenomenon while 
studying evolution of evaporating black holes~\cite{Hawking:1976ra}. 
Suppose we drop some book $A$ into a black hole and observe subsequent 
evolution of the system from a distance.  The book will be absorbed into 
(the horizon of) the black hole, which will then eventually evaporate, 
leaving Hawking radiation.  Now, let us consider another process of 
dropping a different book $B$, instead of $A$, and see what happens. 
The subsequent evolution in this case is similar to the case with $A$. 
In fact, if the masses of $A$ and $B$ are the same, then the masses 
of the black holes after absorbing these books will be the same, so 
the final state radiations obtained after evaporation of these black 
holes are also expected to be the same, because the form of radiation 
depends only on the mass of a black hole in the semi-classical 
approximation~\cite{Hawking:1974sw}, which was believed to be correct 
for large systems like black holes.

\begin{figure}[t]
\begin{center}
  \includegraphics[width=16cm]{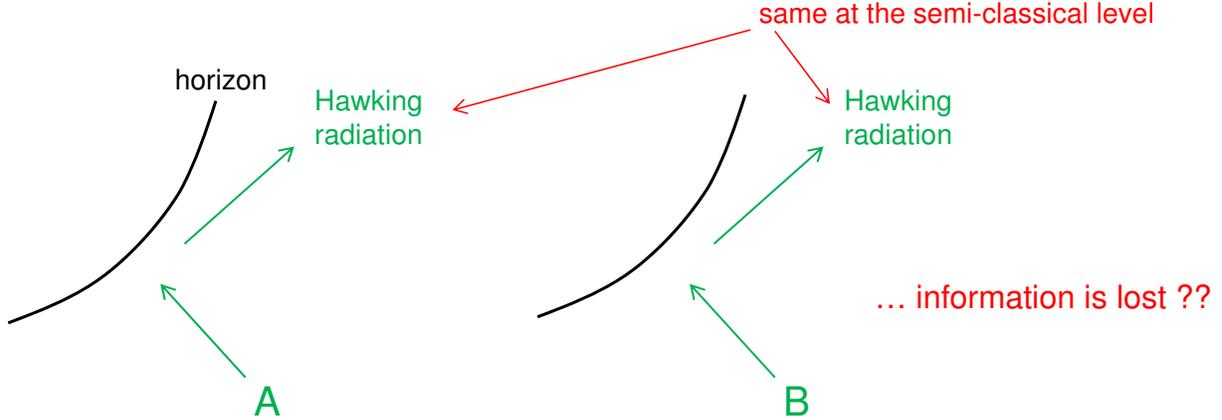}
\caption{If we drop two different objects (e.g.\ $A$ and $B$) into 
 a black hole, the final states seem to be identical at the level of 
 semi-classical approximation, leading to information loss.  This is 
 believed {\it not} to be the case at the full quantum level---final 
 state Hawking radiation contains the full information about the 
 initial state ($A$ or $B$) in the form of subtle quantum correlations 
 between radiation quanta.}
\label{fig:BH1}
\end{center}
\end{figure}
If the final state radiations are really identical---regardless of the 
details of the books---then this would imply that ``information is lost.'' 
Namely, one cannot {\it in principle} recover what was the initial state 
just by looking at the final state of the system; see Fig.~\ref{fig:BH1}. 
(In technical terminology, it is said that unitarity is violated.) 
Who cares?  We care!  In any other situation in physics, we never 
encounter this kind of phenomenon.  For example, Newtonian mechanics 
is deterministic, meaning that if we have perfect knowledge about the 
current state of a system, then we can know its future and past by 
evolving the equation of motion forward and backward in time.  Even 
in quantum mechanics, the Schr\"{o}dinger equation is deterministic, 
so that perfect knowledge of a quantum state should allow us to infer 
its future and past (although, in practice, it is impossible to have 
such knowledge).  To accept the information loss, we need to give up 
usual (unitary) quantum mechanics.

Following recent progress in understanding quantum gravity, 
especially the discovery of the anti de~Sitter/conformal field theory 
duality~\cite{Maldacena:1997re} (which allow us to map certain 
gravitational systems into known, unitary theories), theorists now 
do not think such information loss will actually occur.  We now think 
that the final state radiations obtained from evaporation of the 
black holes that have absorbed book $A$ and $B$ are, in fact, slightly 
different---different in quantum entanglement between many quanta in 
the radiation.  It is simply that when the semi-classical approximation 
is adopted, which discards all the information on such quantum 
correlations, the two final states in Fig.~\ref{fig:BH1} {\it look} 
the same.  This situation is, in fact, not much different from burning 
a book in a (fictitious) classical, Newtonian world.  Even in this case, 
the final states of burning book $A$ and $B$ look quite the same---some 
ashes and dirty air---but if we know the states precisely enough, 
specifically the locations and velocities of all the molecules, then 
the information about the initial states should still be there:\ we 
must be able to solve the Newton equation backward in time to see 
if the initial book was $A$ or $B$.  In this sense, black holes are 
quite ``conventional'' objects---they simply ``burn'' information 
(or ``scramble'' it in technical terms).

\begin{figure}[t]
\begin{center}
  \includegraphics[width=17cm]{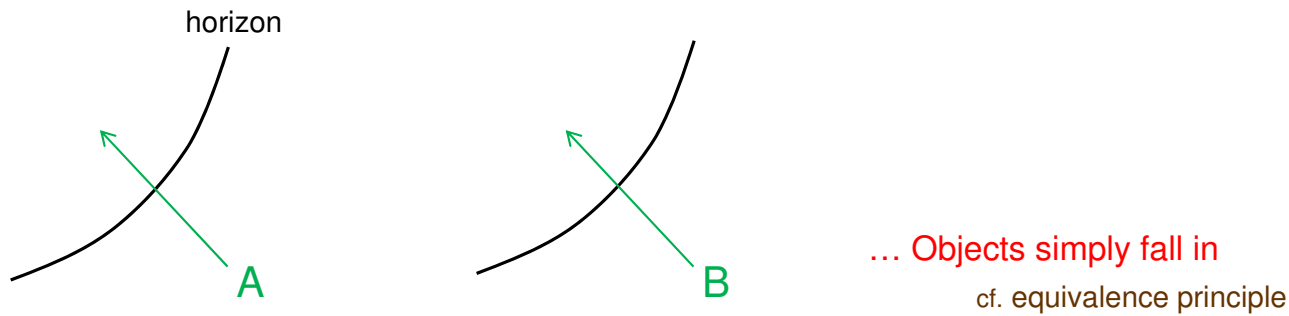}
\caption{If we observe the same process as in Fig.~\ref{fig:BH1} from 
 a falling observer's point of view, the falling object ($A$ or $B$) 
 simply passes the black hole horizon without any disruption.  The 
 full information about the object (in fact, the object itself) will 
 therefore be inside the horizon at late times from this observer's 
 viewpoint.}
\label{fig:BH2}
\end{center}
\end{figure}
A puzzling thing occurs, however, if we observe {\it the same phenomenon} 
from the viewpoint of an observer who is falling into the black hole 
with a book.  In this case, the equivalence principle says that the 
book does not feel gravity (except for the tidal force which is tiny 
for a large black hole), so that it simply passes through the black 
hole horizon without any disruption; see Fig.~\ref{fig:BH2}.  This 
implies that all the information about the book (in fact, the book 
itself) will be {\it inside} the horizon at late times.  On the other 
hand, we have just argued that from a distant observer's point of view, 
the information will be {\it outside}---first on (or more precisely, 
near) the horizon and then in Hawking radiation emitted from the black 
hole.  Which is correct?

One might think that the information is simply duplicated:\ one inside 
and the other outside.  This, however, cannot be the case.  Quantum 
mechanics prohibits faithful copy of full quantum information, due to 
the no-cloning theorem~\cite{Wootters:1982zz}.  (A simple way of seeing 
this is that if a quantum state could be duplicated, then one would be 
able to measure complementary quantities, e.g.\ position and momentum, 
in each copy, contradicting the uncertainty principle.)  Therefore, it 
seems that the two pictures by the two observers cannot both be correct.

The solution to this puzzle is quite interesting---{\it both pictures 
are correct, but not at the same time}.  The point is that one cannot 
be {\it both} a distant observer {\it and} a falling observer at 
the same time.  If you are a distant observer, the system looks 
as in Fig.~\ref{fig:BH1} so that the information will be outside, 
while if you are a falling observer, then the system appears as in 
Fig.~\ref{fig:BH2} and the information (the book itself) will be 
inside.  There is no inconsistency in either picture; only if you 
artificially ``patch'' the two pictures, which you cannot physically 
do, then the apparent inconsistency of information duplication 
occurs. Note that for this argument, the existence of the horizon 
is crucial---because of it, the distant and falling observers cannot 
compare their findings about the location of information, avoiding 
contradiction in a single picture.  This surprising aspect of a system 
with gravity is called black hole complementarity~\cite{Susskind:1993if}.

An important lesson from the above analysis of black holes is that 
quantum states must be defined carefully in a system with gravity. 
From a general relativistic point of view, there is nothing wrong with 
defining quantum states on late-time spacelike hypersurfaces---often 
called nice slices---on which the information exists {\it both} in 
Hawking radiation {\it and} internal space.  This, however, leads to 
(fictitious) duplication of quantum information initially carried by 
a falling object:\ {\it including both Hawking radiation and the inside 
spacetime region within a single description is overcounting}.  To 
avoid this problem, Hilbert space for the quantum states must be 
restricted to the one associated with appropriate spacetime regions:\ 
``your side'' of the horizon.  For example, if you include Hawking 
radiation as well as the horizon degrees of freedom in your description, 
i.e.\ if you are a distant observer, then the internal space of 
the black hole {\it literally does not exist}---including it would 
violate the laws of quantum mechanics.

\subsection{The multiverse as a quantum mechanical universe}
\label{subsec:quantum-univ}

Let us now consider eternally inflating spacetime.  Because of accelerating 
expansion, an inflationary space has a horizon---we cannot see an 
object further than a certain distance, called the de~Sitter horizon, 
because the expansion of space makes any signals from such an object 
unreachable to us.  This situation is simply the ``inside out'' version 
of the black hole case viewed from a distant observer!  In the black 
hole case, we were staying outside the horizon, while now inside; but 
the basic thrust is the same---{\it spacetime on the other side of 
the horizon does not exist}.  Specifically, in an eternally inflating 
spacetime, if you include Gibbons-Hawking radiation (an analogue of 
Hawking radiation in the black hole case), then you should not include 
the region outside the horizon in your description of quantum states. 
More precisely, the horizon here is the stretched apparent horizon, 
which, unlike the event horizon, can be defined locally without knowing 
what happens in the future (and which exists not only in an exponentially 
expanding de~Sitter space but also in other cosmologically relevant 
spacetimes)~\cite{Nomura:2011dt,Nomura:2011rb}.%
\footnote{Possible relevance of black hole complementarity in eternal 
 inflation has been noted earlier in Ref.~\cite{Bousso:2006ev}, although 
 its explicit implementation is different from the one considered here.}

Our current universe is in a phase of accelerating expansion, so 
there is a de~Sitter horizon at about $4.2~{\rm Gpc} \simeq 1.3 \times 
10^{23}~{\rm km}$ away from us.  Its consistent quantum mechanical 
description, therefore, requires us {\it not} to include spacetime 
outside this horizon.  Then what is the multiverse, which we thought 
exists further away beyond the horizon?  The answer is:\ the 
probability!  Given simple initial conditions such as an eternally 
inflating state $\ket{\Sigma}$, the quantum state evolves into a 
superposition of various different cosmic configurations, as shown 
in Eq.~(\ref{eq:multiverse-evolution}).  Each component (or term) 
of the state corresponds to a quantum state on a well-defined 
semi-classical geometry, {\it defined only on your side (which 
we will refer to as inside hereafter) of and on the apparent horizon}. 
In particular, these terms will contain universes like our own, but 
having different $\rho_\Lambda$; and the coefficient of each such 
term $c(\rho_\Lambda)$ will give, after taking into account an 
appropriate weight for ourselves to emerge, the probability density 
of finding a particular value of $\rho_\Lambda$:
\begin{equation}
  P(\rho_\Lambda) \propto |c(\rho_\Lambda)|^2.
\label{eq:P-Lambda}
\end{equation}
(The issue of defining probabilities will be discussed in more 
detail in Section~\ref{subsec:unif}.  For an explicit calculation of 
$P(\rho_\Lambda)$ in the present context, see Ref.~\cite{Larsen:2011mi}.) 
To put it simply, {\it the multiverse lives in probability space}.

Formally, the construction of Hilbert space for the multiverse state 
$\ket{\Psi(t)}$ implementing the picture described above can be made 
quite analogously to the usual Fock space construction in quantum field 
theory~\cite{Nomura:2011dt,Nomura:2011rb}.  For a fixed semi-classical 
geometry ${\cal M}$ (or more precisely, a set of fixed semi-classical 
geometries ${\cal M} = \{ {\cal M}_i \}$ having the same horizon 
$\partial {\cal M}$), the Hilbert space is given by
\begin{equation}
  {\cal H}_{\cal M} = {\cal H}_{{\cal M}, {\rm bulk}} 
    \otimes {\cal H}_{{\cal M}, {\rm horizon}},
\label{eq:ST-H_M}
\end{equation}
where ${\cal H}_{{\cal M}, {\rm bulk}}$ and ${\cal H}_{{\cal M}, {\rm 
horizon}}$ represent Hilbert space factors associated with the degrees 
of freedom inside and on the apparent horizon $\partial {\cal M}$. 
The dimensions of these spaces are given by
\begin{equation}
  {\rm dim}\,{\cal H}_{{\cal M}, {\rm bulk}} 
  = {\rm dim}\,{\cal H}_{{\cal M}, {\rm horizon}} 
  = \exp\left(\frac{{\cal A}_{\partial {\cal M}}}{4}\right),
\label{eq:dim-H_M}
\end{equation}
where ${\cal A}_{\partial {\cal M}}$ is the area of the horizon in units 
of $l_P$.  The fact that the maximum number of degrees of freedom (i.e.\ 
the logarithm of the dimension of the Hilbert space) scales with the 
area, rather than the volume, is a manifestation of the holographic 
principle~\cite{'tHooft:1993gx}, which roughly says that the number 
of degrees of freedom that can be put in a fixed region in a theory 
with general covariance is limited by the area of the surface bounding 
it.  The full Hilbert space for dynamical spacetime is then given 
by the direct sum of the Hilbert spaces for different ${\cal M}$'s
\begin{equation}
  {\cal H} = \bigoplus_{\cal M} {\cal H}_{\cal M},
\label{eq:ST-H}
\end{equation}
which is analogous to the Fock space construction in quantum 
field theory: ${\cal H}_{\rm QFT} = \bigoplus_{n = 0}^{\infty} 
{\cal H}_{\rm 1P}^{\otimes n}$, where ${\cal H}_{\rm 1P}^{\otimes n}$ 
is the $n$-particle Hilbert space.  In addition, the complete Hilbert 
space for quantum gravity must contain ``intrinsically quantum 
mechanical'' states, associated with spacetime singularities:
\begin{equation}
  {\cal H}_{\rm QG} = {\cal H} \oplus {\cal H}_{\rm sing},
\label{eq:QG-H}
\end{equation}
where ${\cal H}_{\rm sing}$ represents the Hilbert space for the 
singularity states.  The evolution of the multiverse state $\ket{\Psi(t)}$ 
is deterministic and unitary in ${\cal H}_{\rm QG}$, but not in 
${\cal H}_{\cal M}$ or ${\cal H}$.

\subsection{``Reference frame dependence'' of the concept of spacetime}
\label{subsec:ref-dep}

In the construction of Hilbert space in the previous section, we 
invoked apparent horizons.  In the cosmological context, however, 
the locations of these horizons are ``observer dependent.''  For example, 
the location of a de~Sitter horizon depends on a spatial point we 
consider to be the center.  What does this really mean?

What we are actually doing here is fixing {\it a reference frame}, 
including ``the origin of the coordinates.''  It is well known that 
to do Hamiltonian quantum mechanics---which we are doing here---we 
must fix all the ``gauge redundancies,'' the redundancies of describing 
the same system in different ways.  A theory of gravity has huge 
redundancies associated with general coordinate transformations, 
and fixing a reference frame (more precisely, electing a local Lorentz 
frame) is precisely a way to eliminate these redundancies and to extract 
physical information, i.e.\ causal relations among events which are 
invariant under general coordinate transformations.  This is so important 
that we repeat it again---{\it we need to fix a reference frame when 
we describe a system with gravity quantum mechanically}.  This makes 
the apparent horizon well-defined:\ it is the horizon as viewed from 
the center (the ``origin'') of the chosen reference frame.  Once a 
reference frame is chosen, the location of a physical object with 
respect to its center has a physical meaning; in particular, spacetime 
outside the horizon does not exist, as we argued earlier.

What happens if we change the reference frame, e.g.\ by a spatial 
translation or boost?  As in any symmetry transformation, this operation 
must be represented by a unitary transformation in the entire Hilbert 
space ${\cal H}_{\rm QG}$.  There is, however, no reason why it 
must be represented in each component ${\cal H}_{\cal M}$.  In 
particular, the transformation can in general mix elements in different 
${\cal H}_{\cal M}$ (as well as those in ${\cal H}_{\rm sing}$). 
Moreover, even if the transformation maps all the elements in 
${\cal H}_{\cal M}$ onto themselves for some ${\cal M}$, there is 
no reason that it should not mix the degrees of freedom associated with 
${\cal H}_{{\cal M}, {\rm bulk}}$ and ${\cal H}_{{\cal M}, {\rm horizon}}$.
\begin{figure}[t]
\begin{center}
  \includegraphics[width=17cm]{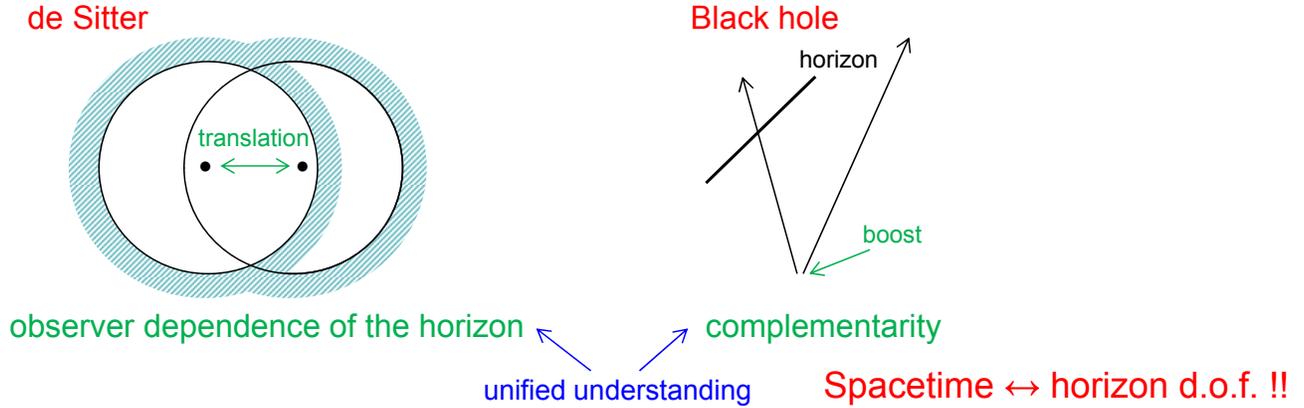}
\caption{If we change a reference frame by a spatial translation in 
 de~Sitter space, what is described as spacetime in one reference frame 
 is described (in part) as the horizon degrees of freedom in the other 
 frame.  In a system with a black hole, a large boost at an early time 
 similarly transforms spacetime (and the singularity) inside the black 
 hole into the horizon degrees of freedom (and Hawking radiation). 
 These phenomena are nothing but the ``observer dependence'' of horizons 
 and black hole complementarity, which can be understood as special 
 cases of the more general transformation associated with reference 
 frame changes.}
\label{fig:ref-change}
\end{center}
\end{figure}

This has a dramatic consequence on the notion of spacetime in a theory 
with gravity~\cite{Nomura:2011rb}.  Consider a state in space with 
accelerating expansion (de~Sitter space).  If we change the reference 
frame by performing a spatial translation, then (a part of) the degrees 
of freedom associated with {\it internal spacetime} in the original 
state must be described as the {\it horizon} degrees of freedom after 
the reference frame change, and vice versa; see Fig.~\ref{fig:ref-change} 
(left).  This implies that what is spacetime and what is not (in this 
case, the horizon) depends on a reference frame!  A more drastic situation 
occurs when there is a black hole.  Consider a reference frame in which 
the center stays outside the horizon at late times.  In this reference 
frame, the internal space of the black hole does not exist, as argued 
before.  Now, let us change a reference frame by performing a boost 
at some early time so that the center of the new reference frame enters 
into the back hole at late times; Fig.~\ref{fig:ref-change} (right). 
In this case what was described as the horizon degrees of freedom (and 
Hawking radiation) in the old reference frame is now described 
as the internal spacetime of the black hole (and the singularity)! 
Note that in either of these examples, before and after the reference 
frame change, we are describing {\it the same (entire) physical system, 
not only their parts}.  It is simply that what is spacetime in one 
reference frame is something else (a horizon, singularity, etc) in 
another---{\it the concept of spacetime depends on the reference frame}.

The phenomena we have just seen are exactly the ``observer dependence'' 
of horizons and black hole complementarity.  They can, therefore, 
be understood in a unified manner as special cases of the general 
transformation (i.e.\ reference frame changes) considered here.  They 
arise because changes of the reference frame are represented in Hilbert 
space ${\cal H}_{\rm QG}$, which contains components ${\cal H}_{\cal M}$ 
that are defined only in restricted spacetime regions because of the 
existence of horizons.

\begin{figure}[t]
\begin{center}
  \includegraphics[width=15cm]{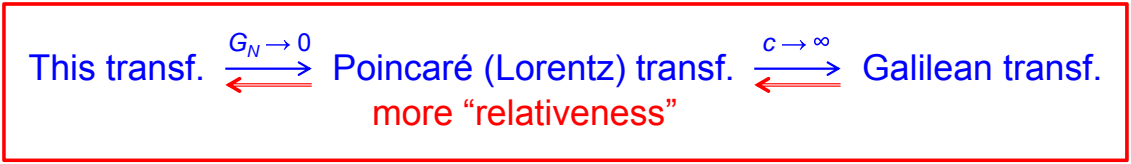}
\caption{The transformation described here is reduced to the Poincar\'{e} 
 transformation of special relativity in the limit $G_N \rightarrow 0$, 
 whose subgroup---the Lorentz transformation---is reduced to the Galilean 
 transformation of Newtonian mechanics in the limit $c \rightarrow \infty$. 
 Physical descriptions of nature become more relative as more fundamental 
 constants of nature are turned on, as represented by the left-pointing 
 arrows.  With gravity ($G_N \neq 0$), a change of the reference frame 
 makes even the concept of spacetime relative.}
\label{fig:transf}
\end{center}
\end{figure}

The transformation described here can be viewed as an extension of 
the Lorentz/Poincar\'{e} transformation in the quantum gravitational 
context~\cite{Nomura:2011rb}; indeed, it is reduced to the standard 
Poincar\'{e} transformation of special relativity in the limit $G_N 
\rightarrow 0$, where $G_N$ is the Newton constant.  This is precisely 
analogous to the fact that the Lorentz transformation (which is a 
subgroup of the Poincar\'{e} transformation) is viewed as an extension 
of the Galilean transformation, which arises as the $c \rightarrow 
\infty$ limit of the Lorentz transformation, where $c$ is the speed 
of light.  In the Galilean transformation a change of the reference 
frame leads only to a constant shift of all the velocities, while in 
the Lorentz transformation it also alters temporal and spatial lengths 
(time dilation and Lorentz contraction) and makes the concept of 
simultaneity relative.  With gravity, a change of the reference frame 
makes even the concept of spacetime relative---general relativity makes 
things really relative in the quantum context!  See Fig.~\ref{fig:transf} 
for the summary of these relations.

\subsection{Unification of the eternally inflating multiverse and many 
 worlds in quantum mechanics}
\label{subsec:unif}

Having defined the multiverse state $\ket{\Psi(t)}$, physical questions 
can now be answered following the rule of quantum mechanics. 
An important point here is that the ``time'' $t$ in quantum gravity is 
simply an auxiliary parameter introduced to describe the ``evolution'' 
of the state, exactly like a variable $t$ used in a parametric 
representation of a curve on a plane, $(x(t), y(t))$.  The physical 
information is only in {\it correlations} between events, like 
correlations between $x$ and $y$ in the case of a curve on a 
plane~\cite{DeWitt:1967yk}.  Specifically, time evolution of a 
physical quantity $X$ is nothing more than a correlation between $X$ 
and a quantity that can play the role of time, such as the location of 
the hands of a clock or the average temperature of CMB in our universe.

Any physical question can then be phrased as:\ given condition $A$ 
we specify, what is the probability for an event $B$ to occur? 
For example, one can specify a certain ``premeasurement'' situation 
$A_{\rm pre}$ (e.g.\ the configuration of an experimental apparatus 
and the state of an experimenter before measurement) as well as a 
``postmeasurement'' situation $A_{\rm post}$ (e.g.\ those after the 
measurement but without specifying outcome) as $A = \{ A_{\rm pre}, 
A_{\rm post} \}$, and then ask the probability of a particular result 
$B$ (specified, e.g., by a physical configuration of the pointer 
of the apparatus in $A_{\rm post}$) to be obtained.  In the context 
of the multiverse, the relevant probability $P(B|A)$ is given 
by~\cite{Nomura:2011dt,Nomura:2011rb}:
\begin{equation}
  P(B|A) = \frac{\int\!\!\!\int\!dt_1 dt_2 \bra{\Psi(0)} U(0,t_1)\, 
      {\cal O}_{A_{\rm pre}}\, U(t_1,t_2)\, {\cal O}_{A_{\rm post} \cap B}\, 
      U(t_2,t_1)\, {\cal O}_{A_{\rm pre}}\, U(t_1,0) \ket{\Psi(0)}}
    {\int\!\!\!\int\!dt_1 dt_2 \bra{\Psi(0)} U(0,t_1)\, 
      {\cal O}_{A_{\rm pre}}\, U(t_1,t_2)\, {\cal O}_{A_{\rm post}}\, 
      U(t_2,t_1)\, {\cal O}_{A_{\rm pre}}\, U(t_1,0) \ket{\Psi(0)}}.
\label{eq:prob-final}
\end{equation}
Here, $U(t_1,t_2) = e^{-iH(t_1-t_2)}$ is the ``time evolution'' operator 
with $H$ being the Hamiltonian for the entire system, and ${\cal O}_X$ 
is the operator projecting onto states consistent with condition $X$. 
Note that since we have already fixed a reference frame, conditions 
$A_{\rm pre}$ and $A_{\rm post}$ in general must involve specifications 
of the locations and velocities of physical objects {\it with respect 
to the origin of the coordinates}, in addition to those of physical 
times made through configurations of non-static objects (e.g.\ the hands 
of a clock or the status of an experimenter).

The integrations over the ``time'' variable $t$ in Eq.~(\ref{eq:prob-final}) 
must be taken from $t = 0$, where the initial condition for $\ket{\Psi(t)}$ 
is specified,%
\footnote{This is the initial condition for a state whose future evolution 
 we want to follow (analogous to the initial condition of a dynamical 
 system that we want to solve in Newtonian mechanics), and {\it not} 
 (necessarily) the initial condition for the entire multiverse; namely, 
 $\ket{\Psi(0)}$ here can simply be a component of the entire multiverse 
 state at some particular moment.  The real ``beginning'' of the quantum 
 universe, i.e.\ the {\it ultimate} boundary condition for the {\it entire} 
 multiverse state, is still an important open issue.  (Note added: for 
 a recent proposal on this issue within the framework discussed here, 
 see Ref.~\cite{Nomura:2012zb}.)}
to $t = \infty$, because conditions $A_{\rm pre}$ and $A_{\rm post}$ 
may be satisfied at any values of $t > 0$ (denoted by $t_1$ and $t_2$ 
in the equation).  Despite the integrals running to $\infty$, the 
formula of Eq.~(\ref{eq:prob-final}) does not involve infinities, which 
would arise if an event occurred infinitely many times with a finite 
probability.  This is because given a generic initial condition, 
the multiverse state $\ket{\Psi(t)}$ at late times will evolve into 
a superposition of terms corresponding to supersymmetric Minkowski 
space (certain highly symmetric space with $\rho_\Lambda = 0$) or 
spacetime singularity:
\begin{equation}
  \ket{\Psi(t)} \,\,\stackrel{t \rightarrow \infty}{\longrightarrow}\,\, 
    \sum_i a_i(t) \ket{\mbox{supersymmetric Minkowski space $i$}}
  \,+\, \sum_j b_j(t) \ket{\mbox{singularity state $j$}},
\label{eq:asympt}
\end{equation}
since these are the only absolutely stable states in the string landscape; 
the coefficients of the other components, including the ones which are 
selected by ${\cal O}_{A_{\rm pre}}$ and ${\cal O}_{A_{\rm post}}$, 
decay exponentially.  This makes the meaning of eternal inflation clear. 
It is the picture obtained by focusing on a component staying in a 
meta-stable de~Sitter state, whose coefficient, however, is decaying 
exponentially with $t$.  In particular, expansion of space does not 
imply the increase of probability.

Equation~(\ref{eq:prob-final}) is our final formula for the probabilities. 
This is essentially the Born rule; indeed, one can show that the formula 
is reduced to the standard Born rule under the usual situation of a 
terrestrial experiment.  There is no freedom of choosing one's own 
(arbitrary) definition of probabilities, and there is no ambiguity 
associated with spatial points separated beyond the horizon, as spacetime 
beyond the horizon simply does not exist.  Furthermore, the formula 
gives a well-defined, finite answer to any physical question we ask. 
Therefore, ...
\begin{equation*}
\ovalbox{The measure problem in eternal inflation is solved.}
\end{equation*}
We emphasize that the uniqueness of the framework (for a given Hilbert 
space, which we take as in Eq.~(\ref{eq:QG-H})) rests crucially on the 
specification of a reference frame, including its origin/center $p$. 
In particular, this requires us to specify ranges of location and 
velocity in which physical objects must lie {\it with respect to $p$}, 
in specifying $A$ and $B$.  This eliminates the ambiguity associated with 
how these objects must be counted.  Of course, there is still a freedom 
of where we put these objects; for example, we could put them at $p$ 
or some other point at rest, or could specify a phase space region 
in which they must be.  But this is the freedom of questions one may 
ask, and not that of the framework itself.  (And the final answer 
does not depend on the location/velocity of reference point $p$, i.e.\ 
the overall relative location/velocity between $p$ and the specified 
configurations in $A$ and $B$, if the multiverse state $\ket{\Psi(t)}$ 
is invariant under the corresponding reference frame changes.)

\begin{figure}[t]
\begin{center}
\begin{picture}(375,200)(-170,0)
  \Line(0,0)(0,36) \Line(0.5,0)(0.5,36) \Line(-0.5,0)(-0.5,36)
  \Line(0,36)(-48,84)
  \Line(-48,84)(-78,102) \Line(-48,84)(-70.8,102) \Line(-48,84)(-60,102)
  \Vertex(-90,126){1} \Vertex(-100,136){1}
  \Vertex(-110,146){1} \Vertex(-120,156){1}
  \Line(0,36)(-24,84) \Text(9.6,72)[]{$\cdots$}
  \Line(-24,84)(-60,132)
  \Line(-60,132)(-84,150) \Line(-60,132)(-78,150) \Line(-60,132)(-68.4,150)
  \Line(-24,84)(-42,132) \Text(-16.8,120)[]{$\cdots$}
  \Line(-24,84)(12,132)
  \Line(12,132)(20.4,150) \Line(12,132)(25.2,150) \Line(12,132)(36,150)
  \Text(7,121)[lt]{\scriptsize different cosmic}
  \Text(12,112)[lt]{\scriptsize histories}
  \Line(-42,132)(-66,180)
  \Line(-66,180)(-78,198) \Line(-66,180)(-74,198) \Line(-66,180)(-67,198) 
  \Line(-42,132)(-50.4,180) \Text(-34.8,168)[]{$\cdots$}
  \Line(-50.4,180)(-58,198) \Line(-50.4,180)(-54,198) \Line(-50.4,180)(-46,198) 
  \Line(-42,132)(-18,180)
  \Line(-18,180)(-17,198) \Line(-18,180)(-13,198) \Line(-18,180)(-6,198) 
  \Text(-21,169)[lt]{\tiny different outcomes}
  \Text(-17,162)[lt]{\tiny for experiments}
  \Line(0,36)(48,84)
  \Line(48,84)(60,102) \Line(48,84)(66,102) \Line(48,84)(78,102)
  \Vertex(85,131){1} \Vertex(95,141){1} \Vertex(105,151){1}
  \Text(35,65)[lt]{\footnotesize different universes}
  \Line(-168,174)(168,174) \Text(176,174)[l]{$\ket{\Psi(t)}$}
  \Line(-90,15)(90,15) \Text(98,15)[l]{$\ket{\Psi(0)}$}
\end{picture}
\caption{A schematic picture for the evolution of the multiverse state 
 $\ket{\Psi(t)}$.  As $t$ increases, $\ket{\Psi(t)}$ evolves into 
 a superposition of states in which various bubble universes nucleate 
 in various spacetime locations.  Each of these states then evolves 
 further into a superposition of states representing various possible 
 cosmic histories, including different outcomes of experiments performed 
 within that universe.}
\label{fig:many-worlds}
\end{center}
\end{figure}
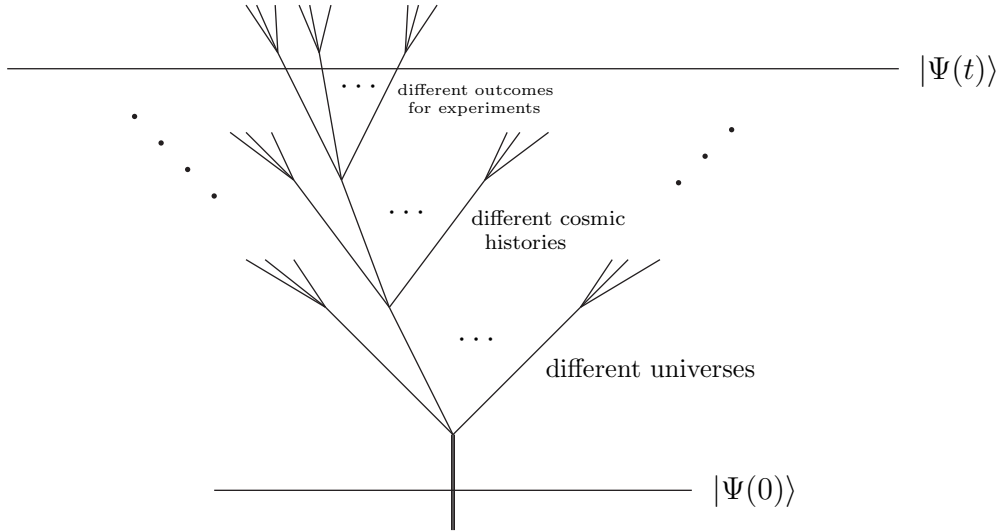

The framework presented here provides a complete account for quantum 
measurement in the multiverse.  Suppose the initial state $\ket{\Psi(0)}$ 
is in an eternally inflating phase.  This state then evolves into a 
superposition of states in which various bubble universes nucleate in 
various spacetime locations.  As $t$ increases, a component representing 
each universe further evolves into a superposition of states representing 
various possible cosmic histories, including different outcomes of 
``experiments'' performed within that universe.  (These ``experiments'' 
may, but need not, be scientific experiments---they can be any physical 
processes.)  For large $t$, the multiverse state $\ket{\Psi(t)}$ 
will therefore contain an enormous number of terms, each of which 
represents a possible world that may arise from $\ket{\Psi(0)}$ 
consistently with the laws of physics.  While evolving, the multiverse 
state experiences both ``branching''~\cite{Everett:1957hd} and 
``amplification''~\cite{q-Darwinism}, responsible, respectively, 
for the realization of various possible outcomes and the appearance 
of a classical world (selecting measurement bases) in each of them. 
Note that since the Hilbert space dimensions of ${\cal H}_{\rm Minkowski}$ 
and ${\cal H}_{\rm sing}$---into which the multiverse state is 
evolving---are infinite, these different worlds do not recohere; they 
really branch into different worlds.  A schematic picture for the evolution 
of the multiverse state is depicted in Fig.~\ref{fig:many-worlds}.

Our probability formula, Eq.~(\ref{eq:prob-final}), can be used to 
answer questions both regarding global properties of the universe and 
outcomes of particular experiments.  For example, given premeasurement 
situation $A_{\rm pre}$, one can ask the probability of finding the 
vacuum energy $\rho_\Lambda$ in a certain range or obtaining a particular 
outcome for an experiment performed in the laboratory, just by adopting 
different conditions $A_{\rm post}$ and $B$.  This, therefore, provides 
complete unification of the eternally inflating multiverse and many 
worlds in quantum mechanics:
\begin{equation*}
\ovalbox{Multiverse = Quantum many worlds.}
\end{equation*}
These two are really the same thing---if one asks a question about a 
global property of the universe, then it is called the multiverse, while 
if one asks a question about outcomes of an experiment/event in our 
everyday life, then it is called quantum many worlds.  They simply 
refer to the same phenomenon occurring at (vastly) different scales.

\section{Conclusions}
\label{sec:concl}

In the past decade or two, a revolutionary change of our view of nature 
has started taking a concrete form---our universe may be one of the 
many in the vast multiverse.  This view is motivated both observationally 
and theoretically:\ the discovery of the nonvanishing vacuum energy 
density in our universe and the string landscape/eternal inflation picture. 
In this article, we have seen that this also comes with a dramatic new 
view of spacetime and gravity, which was forced to resolve the puzzle 
of predictivity crisis that existed in the conventional, semi-classical 
view of eternally inflating spacetime.  We have presented a remarkably 
simple framework that is applicable to physics at all scales:\ from the 
smallest (Planck length) to the largest (multiverse).  We have seen that 
two seemingly different concepts---the multiverse and quantum many 
worlds---are, in fact, the same.  They simply refer to the same phenomenon 
occurring at different length scales.

It is, indeed, quite striking that quantum mechanics does not need any 
modification to be applied to phenomena at such vastly different scales. 
In the 20th century, we have witnessed the tremendous success of quantum 
mechanics, following its birth at the beginning.  In the early 21st century, 
quantum mechanics still seems to be giving us an opportunity to explore deep 
facts about nature, such as spacetime and gravity.  Does quantum mechanics 
break down at some point?  We don't know.  But perhaps, exploring the 
ultimate beginning of the multiverse might provide a key to answer 
that question.

\section*{Acknowledgments}

I would like to thank Alan Guth, Grant Larsen, and Hannes Roberts for 
collaboration and discussion on the subject discussed in this article. 
This work was supported in part by the Director, Office of Science, 
Office of High Energy and Nuclear Physics, of the US Department of 
Energy under Contract DE-AC02-05CH11231, and in part by the National 
Science Foundation under grant PHY-0855653.

\end{document}